[Title] Io's polar volcanic thermal emission indicative of magma ocean and shallow tidal heating models

[Authors]

Ashley Gerard Davies[1,a,], Jason Perry[2], David A. Williams[3], David M. Nelson[3]

1. Jet Propulsion Laboratory – California Institute of Technology, Pasadena, CA 91109, USA.

2. University of Arizona, Tucson, AZ 85721, USA.

3. Arizona State University, Tempe, AZ 85287, USA.

a. Corresponding author - Email: Ashley.Davies@jpl.nasa.gov

[Abstract]

The distribution of Io's volcanic activity likely reflects the position and magnitude of internal tidal heating. We use new observations of Io's polar regions by the *Juno* spacecraft Jovian Infrared Auroral Mapper (JIRAM) to complete near-infrared global coverage, revealing the global distribution and magnitude of thermal emission from Io's currently erupting volcanoes. We show that the distribution of volcanic heat flow from 266 active hot spots is consistent with the presence of a global magma ocean, and/or shallow asthenospheric heating. We find that Io's polar volcanoes are about the same in number per unit area but are less energetic than those at lower latitudes. We also find that volcanic heat flow in the north polar cap is greater than that in the south. The low volcanic advection seen at Io's poles is therefore at odds with measurements of background temperature showing Io's poles are anomalously warm. We suggest that the differences in volcanic thermal emission from Io's poles compared to that at lower latitudes is indicative of lithospheric dichotomies that inhibit volcanic advection towards Io's poles, particularly in the south polar region.

[Text]

The extreme level of volcanic activity on Io, the most volcanically active object in the Solar System[1], is the result of tidally-induced internal heating[2]. Models predict enhanced heat flow at Io's poles if tidal heating is deep in the mantle, and at lower latitudes if heating is predominantly in the asthenosphere, or a magma ocean is present[3-9]. Even although Io's volcanoes have been observed, and thermal emission quantified, at infrared wavelengths for decades[10-15], global mapping of Io's volcanic activity has not been possible until now. Telescopic observations and previous flybys by spacecraft equipped with short-wavelength infrared imagers were mostly confined to view points in the equatorial plane[1] that afforded poor polar coverage. In contrast to these previous observations, when *Juno* entered into a polar orbit of Jupiter in 2016[16], imaging of Io's poles became possible by JIRAM at 3.5 μm and 4.8 μm at spatial resolutions (as of July 2022) of 151 km/pixel to 20 km/pixel.

While numerous hot spots on Io have been previously identified in JIRAM data[16,17], we do not draw any conclusions regarding Io's heat flow from the number of hot spots alone. As the integrated thermal emission from individual hot spots on Io spans more than six orders of magnitude[1,18], estimates of heat flux are more representative of volcanic advection than hot spot number in a given area. A recent study has determined that 4.8-μm spectral radiance is a reasonable proxy for hot spot total thermal emission[15]. While that study used *Galileo* Near Infrared Mapping Spectrometer (NIMS) data obtained from Io observations almost entirely in the equatorial plane (so with poor viewing of the poles), an examination of a selection of Io hot spots observed by JIRAM at both 3.5 μm (L-band) and 4.8 μm (M-band) show that polar hot spots are no different in colour temperature from those at lower latitudes[19]. The implications of this are discussed below.

Here we report the quantification of 4.8-μm spectral radiance from 266 individual hot spots identified in the JIRAM data (Figure 1) obtained between 27 March 2017 (orbit PJ5) and 5 July 2022 (orbit PJ43). This wavelength band is highly sensitive to thermal emission from young lava surfaces on Io between the eruption temperature (~1430 K for basalt) down to ≈200 K[20] (≈70 K above diurnal peak background surface temperature).

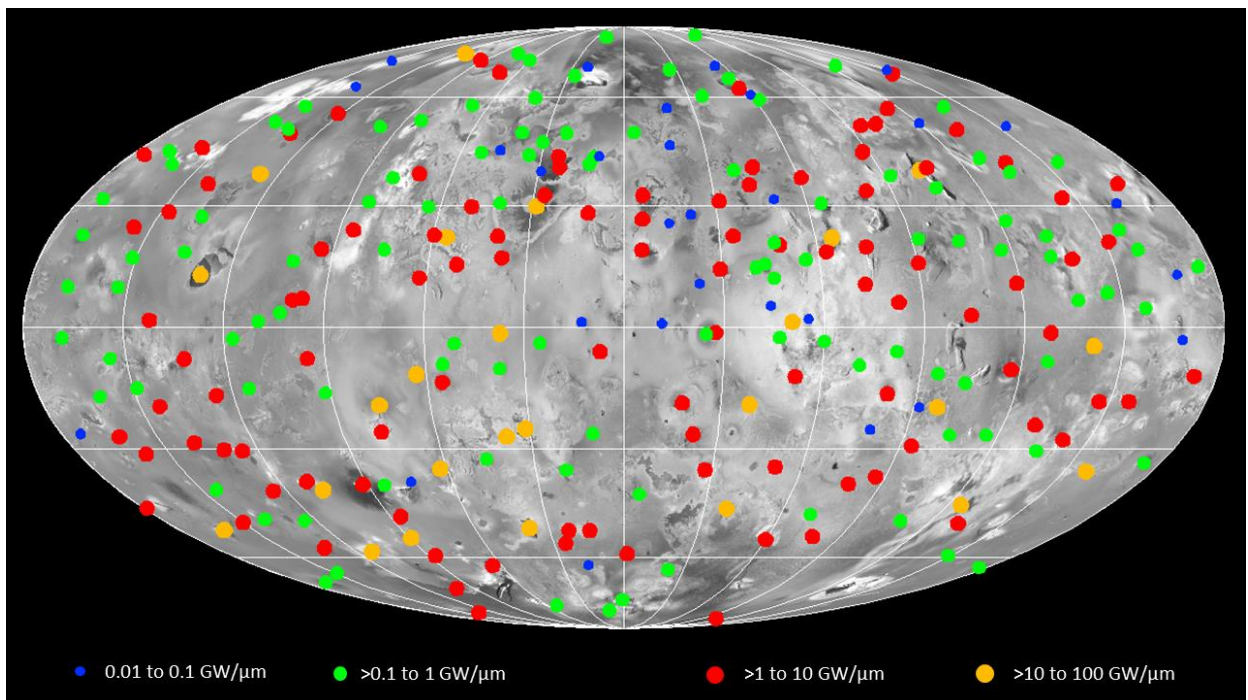

Figure 1. Hot Spot Detections. The maximum, unsaturated M-Band (4.8-μm) spectral radiances from 266 hot spots identified in *Juno* JIRAM data obtained March 2017-July 2022, using data from orbits PJ5 to PJ43 grouped by order of magnitude. The larger the symbol, the greater the 4.8-μm spectral radiance. This is an equal-area Mollweide projection centred on 180 °W, 0 °N. Grid size is 30°. 4.8-μm is a wavelength sensitive to the thermal emission from young, hot lava on Io's surface[20]. Our hot spot detection numbers and distributions (see Tables 1 and 2) differ significantly from those of Zambon et al.[17].

[Results]

We use, in part, superposition of JIRAM images (Figure 2) obtained during each orbital encounter to reveal the presence of low spectral radiance hot spots that otherwise would not be easily seen in single frames. The techniques used to reduce the data are described in the Methodology section. Our set of detected hot spots only overlaps by 64% of those reported from another analysis of the same data[17], but we do find many others (see "Other hot spot detections" in Methodology). This is possibly the result of our geolocation technique creating superposition products with greater positional accuracy and sensitivity to thermal emission from faint hot spots than in the previously cited study[17]. The smallest 4.8-μm spectral radiance we detect is 0.0091 GW/μm in PJ41 data. This radiance is from a patera designated PV32 (JRM183), located at 74.5 °W, 34.5 °N just south of Zal Patera. A table of hot spot locations and maximum unsaturated 4.8-μm spectral radiances, maximum 4.8-μm spectral radiances, and average 4.8-μm spectral radiances, is provided in the Supplementary Table S1. Table 1 summarises our results. We define polar regions as being at latitudes ≥60° because models of interior tidal heating that calculate radially integrated heat flow[3,4] predict enhanced endogenic heat flow above latitude 60° if tidal heating is deep-seated in Io's mantle, and less heat flow at these high latitudes and higher heat flow at lower latitudes if tidal heating is mostly asthenospheric or if a magma ocean is present[9]. We find that polar regions collectively have a slightly lower hot spot density per unit area (hot spots/$10^6$ km$^2$) than those at lower latitudes. However, 4.8-μm spectral radiance density per unit area (GW/μm/km$^2$) from Io's polar caps is less than 50% of that seen at lower latitudes, even when excluding Loki Patera, a powerful thermal source 180 km in diameter located at 310 °W, 12 °N. Loki Patera is a unique feature on Io in terms of size, persistence, and magnitude of thermal emission[1,21]. Previous workers have often treated Loki Patera as an outlier, running two analyses – one including Loki Patera, and one excluding Loki Patera[21]. Excluding Loki Patera makes no significant difference to our conclusions.

Furthermore, there is a difference in hot spot density and spectral radiance density between the polar caps. We find fewer active volcanoes in the south polar cap than in the north. This difference, however, may be the result of south polar observations being fewer and at lower spatial resolution than those of the north polar cap (Figures 2 and 3). The number of hot spots is not as important as how much energy is being delivered to the surface. In terms of 4.8-μm spectral radiance, the south polar volcanoes generate a spectral radiance per unit area of 7 kW/μm/km$^2$, only half that seen in the north (15 kW/μm/km$^2$), and only a quarter of that seen at lower latitudes (27 kW/μm/km$^2$). While higher resolution observations of the south pole may reveal additional hot spots, these are likely to have small outputs as they have so far been undetected. If this is the case, then these hot spots will not significantly alter the south polar value of spectral radiance per unit area. On average, Io's polar volcanoes individually generate less energy than volcanoes at lower latitudes; and the south polar volcanoes generate less energy per volcano than the north polar volcanoes.

In addition to using maximum unsaturated spectral radiance values (Table 1) we also performed the same calculations using saturated pixel values (Table 2) (see Methods). The spectral radiance totals and spectral radiance densities for hot spots where saturated values are used are therefore minimum values. Two south polar cap hot spots (Tiwas Patera – JRM191, and Upulevo Fluctus - JRM214) and two north polar cap hot spots (Tvashtar A – JRM084 and Lei-Kung C – JRM185) have high values due to saturation, but we find that many lower-latitude volcanoes exhibit similarly high saturated radiances so the same conclusions regarding distribution of spectral radiance are drawn.

In further sensitivity analyses we reproduced the hot spot and spectral radiance densities using averaged unsaturated radiance values; and we examined the variability of the spectral radiance quantities as a function of where the polar cap is defined, calculating hot spot spectral radiance density values across a 5°-wide band either side of the 60° latitude boundary. We also ran a model that excluded detections where the emission angle exceeded 75° to ensure that low-latitude volcanoes were not generating higher spectral radiances than polar cap volcanoes simply due to a high emission angle cosine correction.  We find no effect on our conclusions.

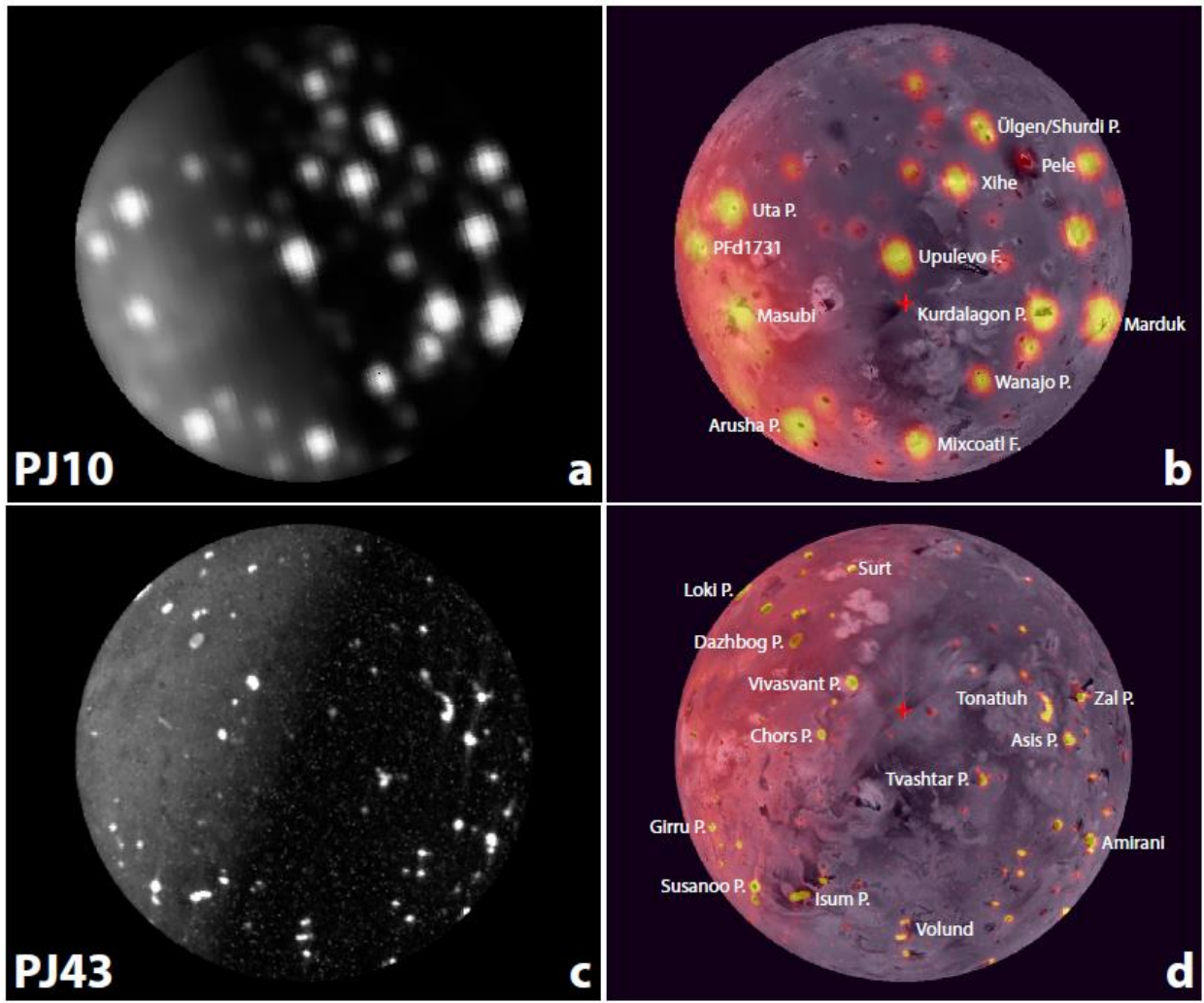

Figure 2. JIRAM 4.8 μm observations of Io. (a) Highest resolution JIRAM observation of Io's south polar region during orbit PJ10 (112 km/pixel) and (b) north polar region during orbit PJ43 (21 km/pixel).  (c) The PJ10 data overlain on the *Galileo/Voyager* base map[37] with some prominent volcanoes identified.  (d) The PJ43 data overlain on the same base map.  Sub-spacecraft points are 317 °W, 78 °S for PJ10 and 175 °W, 79 °N for PJ43.  The red plus is the respective location of the north or south pole.  Differences in appearances between (a) and (b) are due to both the 4.3-fold factor in spatial resolution and differences in exposure time leading to more detector saturation in PJ10 images.

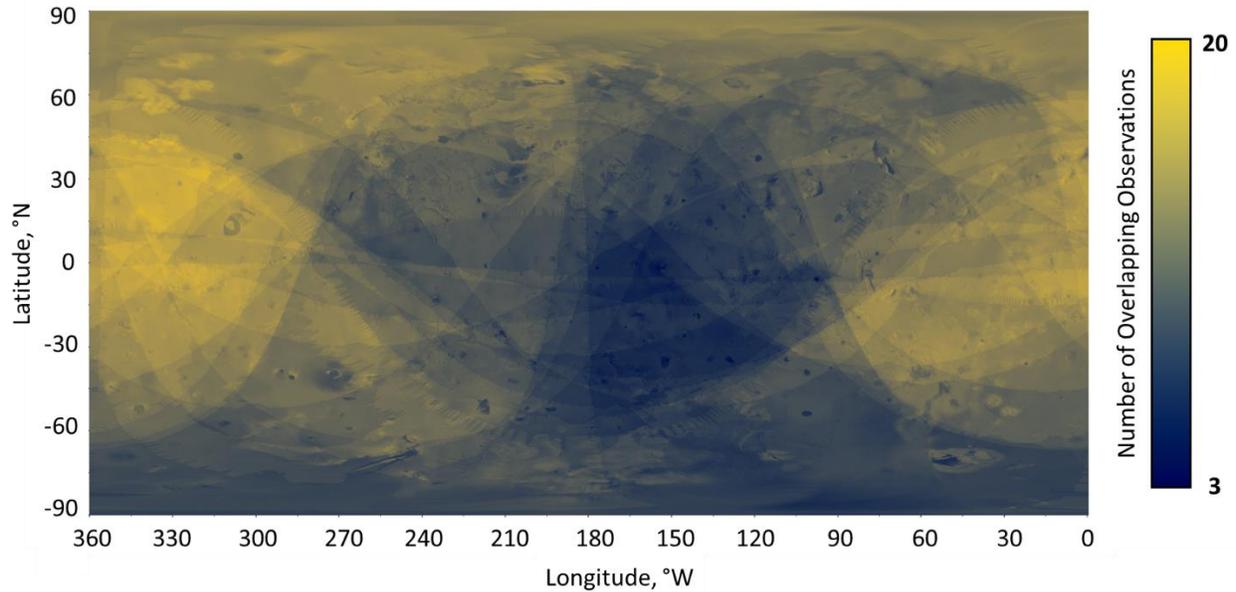

Figure 3. Surface coverage by JIRAM. This figure shows JIRAM surface coverage from orbits PJ5 through PJ43. South polar region coverage is less frequent than that of the north polar region.

[Discussion]

We have created a global map of 4.8-μm spectral radiance that has good sensitivity to high latitudes from 266 volcanic hot spots identified in *Juno* Jovian Infrared Auroral Mapper (JIRAM)[22] data. We find 60% more volcanic spectral radiance per unit area at latitudes lower than 60°. Although the numbers of hot spots per unit area are not greatly different, polar volcanoes emit less energy at 4.8 μm than volcanoes at lower latitudes by a factor of more than two. This result is consistent with models of a global magma ocean or tidal heating in the shallow asthenosphere. Our results also show that from analysis of currently available JIRAM data, north polar volcano spectral radiance is more than twice that of south polar volcanoes, suggesting dichotomies in structure and volcanic advection between polar regions.

Our analysis reveals a global snapshot of the locations where Io's volcanoes are in active eruption, where new lava is currently being emplaced onto the surface as lava flows or exposed in overturning lava lakes[19]. We do not see thermal emission from areas that were active within the last few decades and centuries where surface temperatures have dropped below the JIRAM detection limit, around 180 K. The total 4.8 μm spectral radiance from the 266 hot spots within this dataset is 1.5 TW/μm using maximum measured radiance (including saturated pixels), and 1 TW/μm using maximum unsaturated radiances. Using the empirical relationship between 4.8 μm spectral radiance and inferred total hot spot thermal emission described above[15], these numbers yield estimates of total thermal emission of between 25 and 18 TW. This compares favourably with a previous analysis of 4.8-μm ground-based telescope data which yielded a global volcanic thermal emission estimate of 21 TW[23].

JIRAM so far has therefore detected ~30 TW (≈55%) of the estimated 56 TW emanating from all of Io's volcanic edifices[18,24]. The JIRAM-derived volcanic thermal emission estimate is ≈28% of Io's total global thermal emission (≈106 TW)[25].

Volcanic 4.8-μm spectral radiance has been shown to be (and used as) a proxy for total hot spot thermal emission from the analysis of *Galileo* Near Infrared Mapping Spectrometer (NIMS) data[15,26], and this study shows that Io's polar volcanoes are not particularly different from those at lower latitudes in all aspects except areal extent. If Io were being predominantly tidally heated in the deep mantle, then the expectation is that Io's polar volcanoes would reflect this by being more numerous, or larger and more energetic, and having higher temperature lavas[27]. Regarding the first term, we find only slightly fewer hot spots in polar regions. Nor do we find that the detected hot spots are larger in terms of power output than those at lower latitudes to make up for any scarcity at higher latitudes.

The two latter terms – energy and temperature - are harder to disentangle. Considering JIRAM imager wavelengths, it is important not to interpret a higher colour temperature as evidence of higher magma eruption temperature. JIRAM imager data cannot be used to tightly constrain lava eruption temperature (and hence lava composition) from measurements of thermal emission. Such a constraint can only be imposed using a shorter wavelength (0.4 to 1.5 μm) imager[28]. The shape of the near infrared thermal emission spectrum is mostly determined by the manner in which lava is erupted; a lava fountain has a high colour temperature (which is itself an approximation of a complex surface temperature-area distribution), while a lava flow of the same composition lava with a well-developed surface crust has a much lower colour temperature. Nevertheless, we do not find evidence of higher colour temperatures at polar-region volcanoes, suggesting eruption styles are not greatly dissimilar to those at lower latitudes. It is likely that, prior to the *Juno* encounters, analyses examining high latitude eruptions were skewed by poor spatial resolution and restricted viewing angles, and where detections were subsequently dominated by vigorous eruption styles, such as lava fountaining, that were highly energetic and which generated higher than average colour temperatures, while smaller, more numerous eruptions were not detected.

An open question, however, is whether the distribution and magnitude of volcanic thermal emission itself is a proxy for global heat flow[24]. Io's polar temperatures appear to be warmer than accounted for by solar insolation[29] and it is not known if this results from high endogenic polar heat or from some property of polar surface material with, for example, a high thermal inertia[24,29]. If high endogenic heat flow is the cause, it is also possible that this could promote more polar volcanism. Yet Io's polar volcanoes are less powerful and slightly fewer per unit area than those at lower latitudes. It may be that this apparent conundrum will not be resolved until a spacecraft maps Io's global background endogenic heat flow pattern, separate from heat flow at the ≈2% of Io's surface occupied by active volcanoes[21,30].

The principal means of testing models of Io's interior heating and volcanic advection comes from observing the distribution and magnitude of volcanic activity, thermal emission, and lava composition[27]. The coupling of Io to Europa and Ganymede via the orbital resonance[2] plays an important role in how the strength of the tidal dissipation (and by implication, interior state) has evolved. If Io oscillates into and out of the orbital resonance[31], this has strong implications for Io's volcanism. When Io's orbital eccentricity is damped, tidal heating decreases and deep mantle heating decreases. A lack of enhanced polar volcanism suggests that solid body dissipation in Io's deep mantle is not taking place[27]. In addition to mapping Io's background thermal emission, constraining Io's interior state will likely depend on future spacecraft measurements of the strength of magnetic induction, having separated Io's magnetic induction from effects caused by the Io plasma torus[32,33], coupled to measurements of Io's libration and diurnal tidal potential Love number, $k_2$[34]. Excluding the deep mantle heating model as a consequence of this analysis of JIRAM data means that a strong magnetic induction implies a magma ocean is present, which would be supported by a high value of $k_2$ and large libration[34]; whereas a weak magnetic induction and low $k_2$ and libration values imply solid body dissipation in the asthenosphere[27,34,35].

As *Juno* obtains even closer observations of Io, the quality of thermal detections improves with increasing spatial resolution. We have seen how higher resolution images of individual hot spots show more thermal structure. A good example is Tvashtar Paterae. JIRAM observations obtained during PJ43 (5 July 2022) show multiple hot spots within the two northern paterae of the Tvashtar Paterae complex (Figure 4). For the purposes of quantifying thermal emission by volcano location, care is needed to determine if these hot spots are separate eruptions or different parts of a single event. Another example is Prometheus, which appears as a single source in data up to PJ43, when two hot spots are resolved in higher-resolution images.

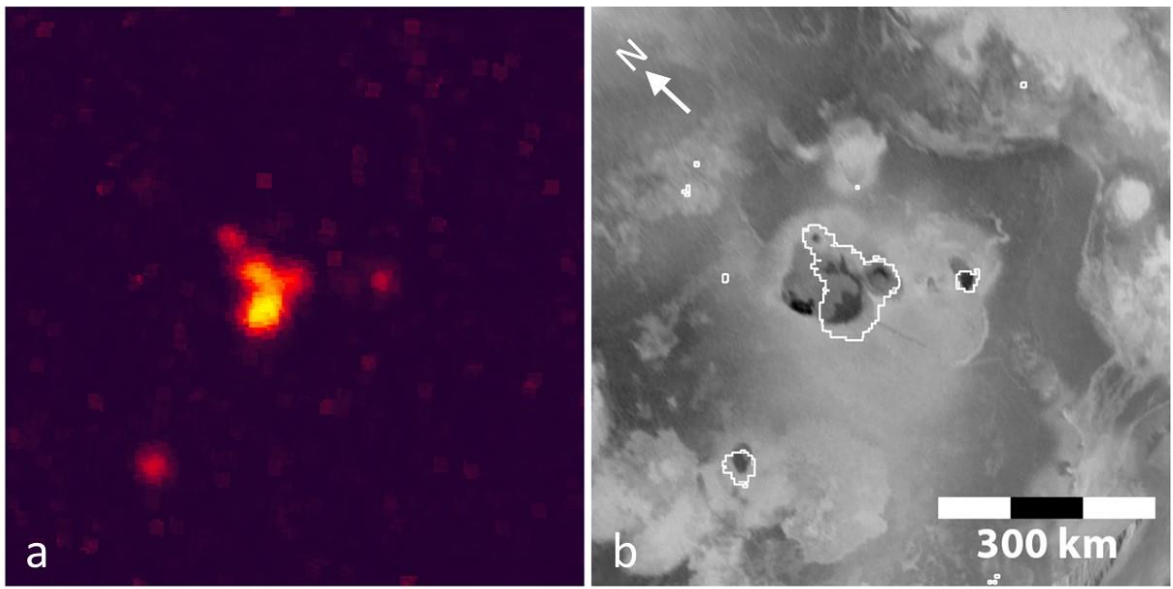

Figure 4. Extensive volcanic activity at Tvashtar Paterae. (a) JIRAM super-resolution image coverage of Tvashtar Paterae (centre of image) and the surrounding region from PJ43 (5 July 2022). These images are centred on 124.3 °W, 62.3 °N. At 20 km/pixel, details of the surface distribution of thermal emission at Tvashtar Paterae become apparent. (b) shows the outlines (solid white line) of the prominent thermal sources overlain on the Io global mosaic[37]. Tvashtar Paterae is the site of episodic, vigorous volcanic activity, including lava fountaining feeding the emplacement of lava flows and possibly lava lake overturn[38,39].

[Corresponding Author]

All correspondence should be addressed to Ashley Davies (Ashley.Davies@jpl.nasa.gov).

[Methods]

Data reduction.  JIRAM data are processed, pixel by pixel, from band radiance (W/m$^2$/str) to spectral radiance (W/m$^2$/str/μm) and corrected for distance from spacecraft to surface (i.e., spacecraft altitude). Data are further corrected for emission angle, assuming Lambertian emission, to calculate surface leaving radiance in W/μm (reported in units of GW/μm for convenience).  The M-band (4.8-μm) filter band width is 0.4975 μm[16].   The JIRAM instantaneous field of view (IFOV) = $\Omega$ = 2.37767 x 10$^{-4}$ rad.   For each pixel, therefore, the following steps are taken to determine surface leaving spectral radiance.   An emissivity ε of 1 is used.

1.  Calculate pixel spatial resolution = $R_{spatial}$ = $\Omega$ $S$  where $S$ = spacecraft altitude (km).

2.  Calculate pixel area $A_{pix}$ (km$^2$) = $(\Omega\ S)^2$.

3.  Convert band radiance $I_{band, \lambda}$ to spectral radiance $I_{spectral, \lambda}$ (W/m$^2$/str/μm) by dividing $I_{band, \lambda}$ by filter width.

4.  Multiply $I_{spectral, \lambda}$ by pixel area in m$^2$ to convert to W/str/μm.

5.  Multiply $I_{spectral, \lambda}$ by $\pi$ to convert to W/μm, assuming Lambertian emission.

6.  An emission angle $e$ correction yields the surface-leaving spectral radiance in W/μm = $I_{spectral, \lambda}$ / cos($e$).

7.  The spectral radiance is summed for the pixels comprising the hot spot.

Frame registration.  We initially investigated hot spot detection in image products navigated solely using NAIF SPICE kernels. We found that the products required additional positional correction.  Using a superposition technique (that allows identification of faint thermal sources that otherwise would be difficult to spot in individual JIRAM frames.

1.  Using *Juno* mission NAIF SPICE kernels, Io is identified within each JIRAM image frame.

2.  Positions are adjusted using limb-fitting and tying hot spots to known stable surface features (the former if lighting conditions allow).  This is a vital step in identifying surface locations as SPICE kernels alone have insufficient precision. These adjustments improve the alignment of hot spots from image to image; reduce motion blur in the resulting summed images; and make it easier to map hot spots to known surface features.   For example, we incorporated 12 images into the PJ17 superposition product.  The range of adjustments needed to align all images was -1.25 pixels in the x direction and 0.75 to 2.75 pixels in the y direction.  As the sub-spacecraft spatial resolution was ≈68 km/pixel (from a spacecraft altitude of ≈289,000 km), these adjustments are large distances on Io's surface, translating to large differences in longitudinal position estimation at high latitude.

3.  For each on-moon pixel, pixel centre latitude, longitude, spacecraft altitude, and emission angle are calculated.  We create saturation masks for each JIRAM observation based on detector sensitivity and spectral radiance for both L- and M-band images so saturated pixels are flagged.

4.  Each Io-containing frame in an orbit is reprojected to the same point perspective map projection at 10x spatial scale of the original data (5x for PJ41).

5.  A superposition product is generated from the average of each of these frames for each orbit.

6.  We examine each detection in each superposition image, being careful to avoid counting "detections" caused by data artifacts (such as diffraction spikes).  Latitude, longitude, size (number of pixels), and average band radiance are measured for each detection; using emission angle and altitude, spectral radiance (GW/μm) is calculated for each hot spot.

7.  For hot spots where radiance is detected in pixels and/or portions of pixels off Io's disc, additional super-position images were generated with three-pixel offsets to the X or Y adjustments used in step 2.  This allows the total band radiance to be measured and the spectral radiance to be calculated for these high-emission angle hotspots.

This methodology prevents misidentification of hot spots and results in a robust dataset.  Detections are shown in Figure 1 and listed in the Supplementary Table S1.

Background removal.  For hot spots identified in daylight, the reflected sunlight component is removed by subtracting the average radiance value of the pixels surrounding the hot spot from each hot pixel in the detection, a technique previously described and applied to NIMS data[26].

Hot spot detection.  No single threshold value of radiance was used to detect hot spots due to differences in background noise and lighting condition, which could make detection more difficult when looking for fainter spots. In nightside/eclipse observations, a threshold brightest pixel >50% greater than background level was used.  For hot spot detections on the dayside, the threshold is lower, sometimes as low as 1.2x background level.  Additional criteria were applied to the faintest dayside detections.  Was the spot seen as a hot spot at other times on the nightside or in eclipse?  Is the proposed hot spot in an area of high albedo, such as Ra Patera and Acala Fluctus?  If so, and other criteria are not met, then no detection is recorded.  By adopting a conservative approach, some very faint hot spot candidates are not included in the detections list.  This approach does not adversely affect the conclusions of the paper as these hot spots are at the low end of the power spectrum and therefore contribute little to total 4.8-μm spectral radiance.

Other hot spot detections.  We have compared our analysis with previous hot spot identifications by Zambon et al.  derived from the same JIRAM dataset[17].  Zambon et al. used hot spot count alone to conclude that Io's polar regions were more volcanically active than at lower latitudes, the precise opposite of our findings.  Possibly explaining this discrepancy, we find significant differences between the numbers and positions of our hot spot detections and the Zambon et al. hot spot list.  Over orbits PJ10 to PJ33 we identify 240 hot spots.  Of the 243 hot spots listed by Zambon et al. (2023) we find 156 (Supplementary Figure S1).  Supplementary Figure S2 shows the comparison of analyses of PJ17 data.  We surmise that many hot spots detected by Zambon et al. are multiple detections of mostly high latitude hot spots, and the superposition technique we use is more sensitive to fainter hot spots not detected by Zambon et al. (2023).

Uncertainty analysis.  JIRAM band radiance data are not provided with uncertainties.  To calculate uncertainties in the spectral radiance density value, the mean, standard deviation, and skewness of spectral radiance are calculated.  These values are reported in Tables 1 and 2.  Given that this non-negative set of data produces standard deviations greater than half of the mean values indicates that the data deviates substantially from a bell shaped (normal) curve. This is indeed the case, reflected in the large positive values of skewness. Rather than use the mean value and standard deviation to quantify uncertainty, the standard deviation and skewness are used to calculate the Standard Error, also reported in Tables 1 and 2. A value of twice the Standard Error is used to represent uncertainty in spectral radiance density in Tables 1 and 2.

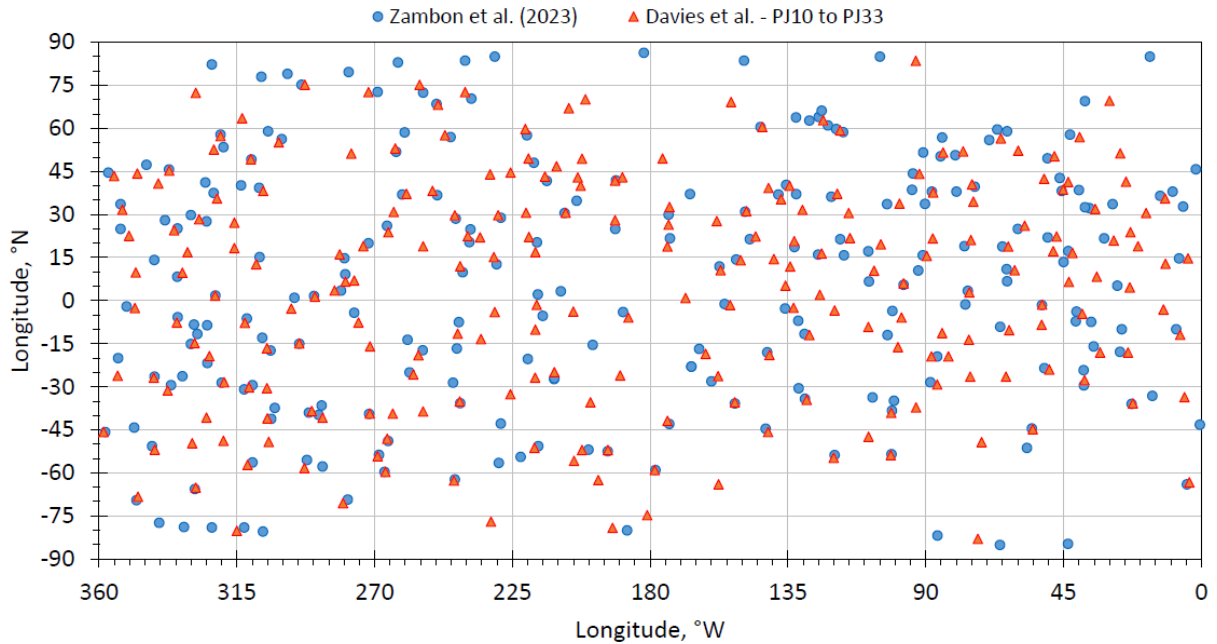

Figure S1. Comparison of hot spot detections in Zambon et al. (2023) and this analysis, limited to orbits PJ10 through PJ33. We find fewer hot spots in polar regions than Zambon et al. and more hot spots at lower latitudes. The discrepancies between analyses increase towards high latitudes, in particular above 75°.

Saturated data. JIRAM band radiance values plateau at the point of detector saturation. The JIRAM detector saturates in a predictable manner. Supplementary Figure S3 shows the saturation band radiance as a function of image integration time. We identify and flag all saturated pixels and have created a saturation mask for every frame in every JIRAM observation of Io.

Data availability. The data analysed during this study are available from NASA's Planetary Data System (PDS) at the URL below. The products generated by this analysis will be available via the Io Geographical Information System Database[36] at Arizona State University on publication.

[https://pds-atmospheres.nmsu.edu/data_and_services/atmospheres_data/JUNO/jiram_orbits.html](https://pds-atmospheres.nmsu.edu/data_and_services/atmospheres_data/JUNO/jiram_orbits.html)

The Io Global Photomosaic[34] used as a background image in Figures 1, 2, 4, and Supplementary Figure S3 is available from https://www.usgs.gov/media/images/io-global-image-mosaic-and-geologic-map

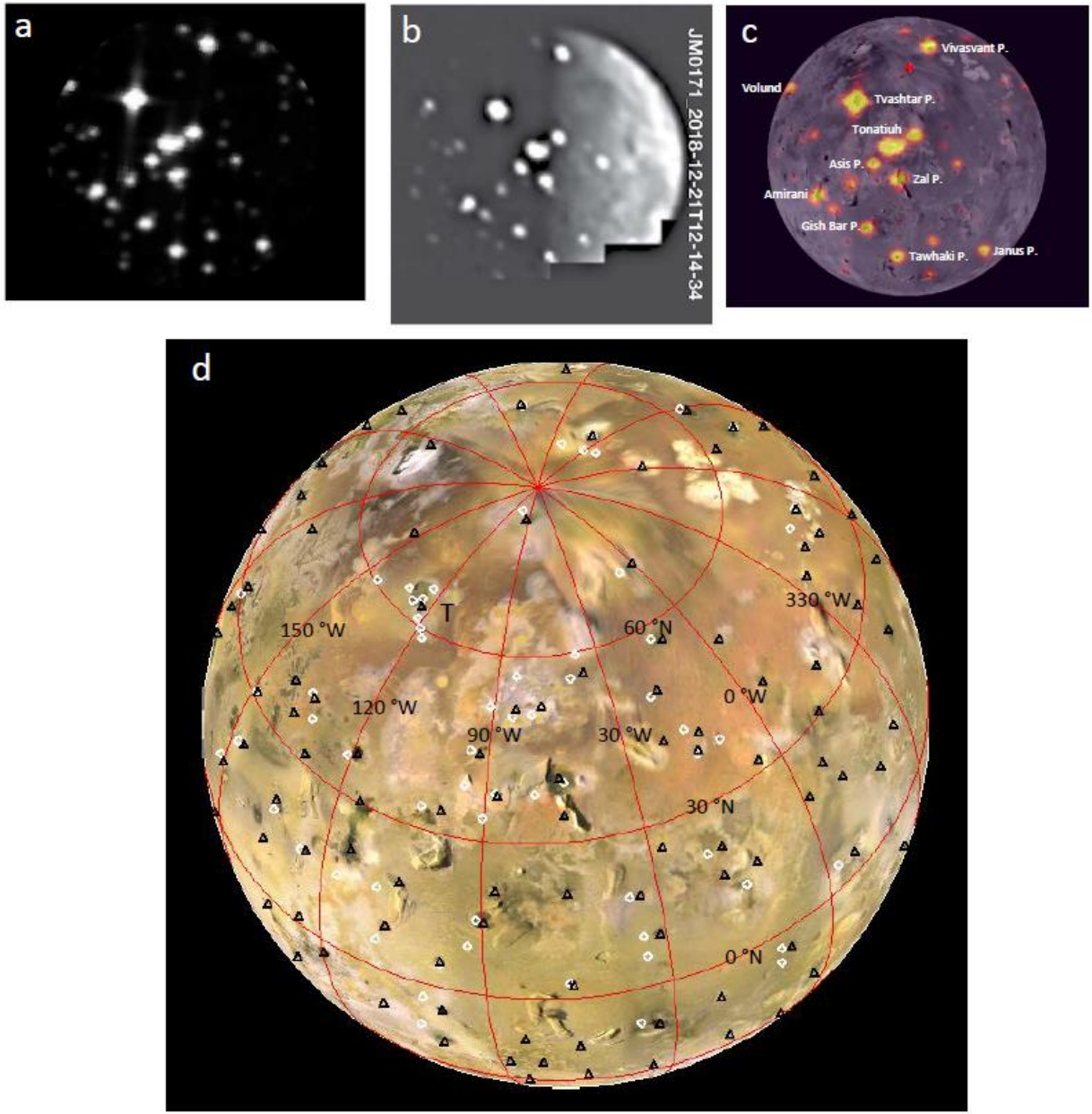

Figure S2.  Comparison of superposition images for PJ17 (21 December 2018) from (a) this analysis and (b) Zambon et al. (2023).  Main hot spots found in this analysis are shown in (c), where the north pole is marked with a red cross.  Io in panels a-d is shown with the same viewing geometry and orientation.  (d) shows hot spots identified in PJ17 data in this analysis (black triangles) compared with hot spots listed by Zambon et al. (white triangles) in PJ17 data.  Data are plotted on the *Voyager-Galileo* mosaic[37].  We identify 103 hot spots.  Zambon et al. identify 64.  At latitudes above 60 °N, we find 8 hot spots, but Zambon et al. find 13.  At Tvashtar Paterae (a group of calderas centred on ≈119 °W, ≈60 °N, designated by "T" in panel d), we identify a single thermal source.  Zambon et al. identify seven, with an eighth close by that we do not detect.

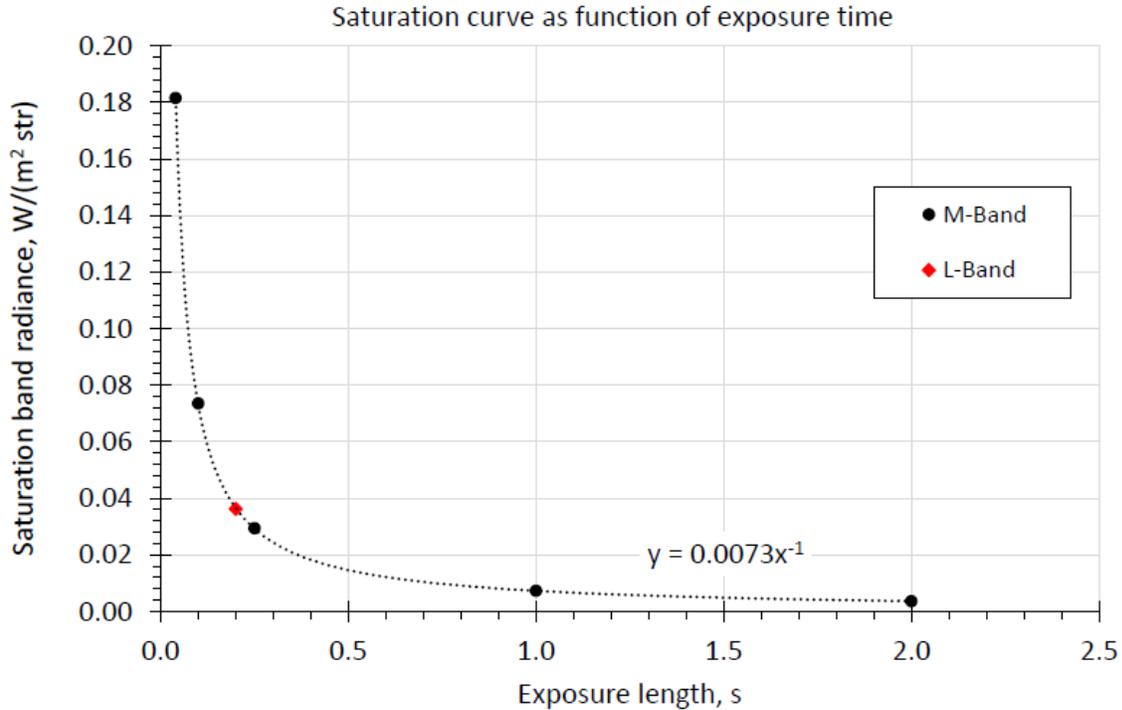

Figure S3. JIRAM detector saturation limit in units of band radiance as a function of exposure duration.

Saturated data. JIRAM band radiance values plateau at the point of detector saturation. The JIRAM detector saturates in a predictable manner. Supplementary Figure S3 shows the saturation band radiance as a function of image integration time. We identify and flag all saturated pixels and have created a saturation mask for every frame in every JIRAM observation of Io.

Data availability. The data analysed during this study are available from NASA's Planetary Data System (PDS) at the URL below. The products generated by this analysis will be available via the Io Geographical Information System Database[36] at Arizona State University on publication.

https://pds-atmospheres.nmsu.edu/data_and_services/atmospheres_data/JUNO/jiram_orbits.html

The Io Global Photomosaic[34] used as a background image in Figures 1, 2, 4, and Supplementary Figure S3 is available from https://www.usgs.gov/media/images/io-global-image-mosaic-and-geologic-map

Code availability. We include python scripts for data projection, and for the generation of data backplanes which for all pixels, if the pixel is on planet, corrected latitude and longitude, band radiance, emission angle, spacecraft altitude, a saturation flag, and spectral radiance.

https://github.com/volcanopele/juno

[Acknowledgements]

This work was performed in part at the Jet Propulsion Laboratory–California Institute of Technology, under contract to NASA, supported by NASA NFDAP award 80NM0018F0612 to AGD, JP, DAW, and DN. We thank Lionel Wilson and Alfred McEwen for comments on a pre-submission version of the manuscript. © 2023. All rights reserved.

[Author contributions]

AGD and DAW organised the research project. AGD administered the project as Principal Investigator. JP, AGD and DN created the data processing pipeline. JP performed primary data processing. AGD and DN provided the second stage of data processing. AGD, JP, and DAW analysed the results. All authors discussed the results and commented on the paper.

[Competing Interests Statement]

The authors declare no competing interests.

[Tables]

| Region of Io | Latitude range | Area | No of hot spots | Hot spot density | Total 4.8-µm spectral radiance | Mean hot spot 4.8-µm spectral radiance | Standard deviation 4.8-µm spectral radiance | Skewness in 4.8-µm spectral radiance | Standard error in 4.8-µm spectral radiance | 4.8-µm spectral radiance density |
|---|---|---|---|---|---|---|---|---|---|---|
| | ° | km$^2$ | | number/10$^6$ km$^2$ | GW/µm | GW/µm | GW/µm | GW/µm | GW/µm | kW/µm/km$^2$ |
| North polar cap | 60 to 90 | 2.77E+06 | 20 | 7.21 | 40.79 | 2.04 | 5.03 | 3.17 | 1.12 | 14.71 ± 5.32 |
| South polar cap | -60 to -90 | 2.77E+06 | 12 | 4.33 | 19.80 | 1.65 | 2.03 | 1.56 | 0.59 | 7.14 ± 2.58 |
| Both polar caps | 60 to 90 and -60 to -90 | 5.54E+06 | 32 | 5.78 | 60.59 | 1.89 | 4.12 | 4.22 | 0.33 | 10.94 ± 1.97 |
| Lower latitudes | < 60 and > -60 | 3.59E+07 | 234 | 6.52 | 946.43 | 4.04 | 8.08 | 4.43 | 0.53 | 26.39 ± 0.74 |
| Lower latitudes, ex. Loki Patera | < 60 and > -60 | 3.59E+07 | 233 | 6.49 | 877.91 | 3.77 | 6.65 | 3.74 | 0.45 | 24.45 ± 0.68 |
| Global | 90 to -90 | 4.14E+07 | 266 | 6.43 | 1006.83 | 3.80 | 7.74 | 4.30 | 0.47 | 24.32 ± 0.59 |

Table 1. Distribution of hot spots and 4.8-µm spectral radiance (orbits PJ5 to PJ43) using maximum unsaturated values and using a 60° latitude cap

| Table 2. Distribution of hot spots and 4.8-µm spectral radiance (orbits PJ5 to PJ43) including saturated values and using a 60° latitude cap | | | | | | | | | | |
|---|---|---|---|---|---|---|---|---|---|---|
| Region of Io | Latitude range | Area | No of hot spots | Hot spot density | Total 4.8-µm spectral radiance | Mean hot spot 4.8-µm spectral radiance | Standard deviation 4.8-µm spectral radiance | Skewness in 4.8-µm spectral radiance | Standard error in 4.8-µm spectral radiance | 4.8-µm spectral radiance density |
| | ° | km² | | number/10⁶ km² | GW/µm | GW/µm | GW/µm | GW/µm | GW/µm | kW/µm/km² |
| North polar cap | 60 to 90 | 2.77E+06 | 20 | 7.21 | 58.67 | 2.93 | 6.66 | 3.10 | 1.49 | 21.18 + 7.65 |
| South polar cap | -60 to -90 | 2.77E+06 | 12 | 4.33 | 44.52 | 3.71 | 6.37 | 2.32 | 1.84 | 16.07 ± 5.80 |
| Both polar caps | 60 to 90 and -60 to -90 | 5.54E+06 | 32 | 5.78 | 103.19 | 3.22 | 6.46 | 2.43 | 1.02 | 18.63 ± 3.36 |
| Lower latitudes | < 60 and > -60 | 3.59E+07 | 234 | 6.52 | 1388.09 | 5.93 | 15.69 | 7.13 | 1.03 | 38.71 ± 1.08 |
| Lower latitudes, ex. Loki Patera | < 60 and > -60 | 3.59E+07 | 233 | 6.49 | 1233.96 | 5.30 | 11.79 | 7.35 | 0.81 | 34.37 ± 0.96 |
| Global | 90 to -90 | 4.14E+07 | 266 | 6.43 | 1494.39 | 5.62 | 14.90 | 7.12 | 0.91 | 36.02 ± 0.87 |